\documentclass{PoS}
\usepackage{amsfonts,amsmath}
\usepackage{graphicx}
\usepackage{amsmath,amssymb,fancyhdr}

\newcommand{\be}{\begin{eqnarray}}
\newcommand{\ee}{\end{eqnarray}}
\newcommand{\nn}{\nonumber}
\newcommand{\nl}{\nonumber \\}

\title{Non-$\gamma_{5}$hermiticity minimal doubling fermion}

\ShortTitle{Non-$\gamma_{5}$hermiticity minimal doubling fermion}

\author{\speaker{Syo Kamata}\\
         
         Department of Physics, Rikkyo University, Tokyo 171-8501, Japan\\
         E-mail: \email{skamata@rikkyo.ac.jp}}

\author{Hidekazu Tanaka\\
  Department of Physics, Rikkyo University, Tokyo 171-8501, Japan\\
  E-mail: \email{tanakah@rikkyo.ac.jp}}

\abstract{We formulate new two dimensional fermions breaking $\gamma_{5}$hermiticity, based on the minimal doubling fermion.
We investigate their properties: (I) Symmetries, (II) eigenvalue distributions, and (III) the number of poles.
As a simple application of the fermions, the Gross-Neveu model in two dimensions is studied using the fermion.
We obtain the parity broken phase diagram called Aoki phase and the chiral broken phase diagram of the model with an imaginary chemical potential.}

\FullConference{31st International Symposium on Lattice Field Theory - LATTICE 2013\\
July 29 - August 3, 2013\\
Mainz, Germany}

\begin{document}

\section{Motivation}

In lattice gauge theory, we often encounter situations that $\gamma_{5}$hermiticity cannot be preserved, e.g., finite density theory, supersymmetric theory, and so on.
Since $\gamma_{5}$hermiticity is needed for reality of Dirac determinant, most of the observables are complex numbers in the case.
In $\gamma_{5}$hermiticity broken theory, such as finite density theory, the partition function does not converge, so that the analysis of high density region using lattice simulation breaks down even if the rewighting method are employed. 
The serious problem is called the sign problem.
Additionally, in odd dimensions, $\gamma_{5}$hermiticity cannot be defined because we cannot define an irreducible matrix which anti-commutes with all spatial $\gamma$-matrices.
For the reasons mentioned above, it is important to know the effects of $\gamma_{5}$hermiticity breaking.
In this work, we consider some effects of $\gamma_{5}$hermiticity breaking.
To examine the effects, we firstly formulate two dimensional fermions without $\gamma_{5}$hermiticity, based on the minimal doubling fermion \cite{Karsten:1981gd, Creutz:2007af}, and then investigate some properties of the fermions. 
We also apply the fermion to the Gross-Neveu model \cite{Gross:1974jv} and obtain parity broken and chiral broken phase diagrams.

\if
This paper is organized as follows.
In section \ref{sec:non-gamma5}, we define non-$\gamma_{5}$hermiticity fermion based on the minimal doubling fermion and investigate their symmetries and properties.
In section \ref{sec:aoki}, we study parity broken phase diagrams for the two dimensional Gross-Neveu model with or without an imaginary chemical potential using the non-$\gamma_{5}$hermiticity fermion.
Final section is devoted to the summary.
\fi

\section{Two dimensional non-$\gamma_{5}$hermiticity fermion} \label{sec:non-gamma5}
Firstly, we define two dimensional fermions without $\gamma_{5}$hermiticity (non-$\gamma_{5}$hermiticity fermion), based on the minimal doubling fermion which preserves exact chiral symmetry \cite{Karsten:1981gd, Creutz:2007af},
\be
D_{1}^{(2)}(p) &=&  \sum_{\mu=1,4} i \ \sin p_{\mu} \cdot \gamma_{\mu} + \kappa ( 1-\cos p_{1}  ) \cdot \gamma_{1} ,\label{eq:mod2-1p}\\
D_{2}^{(2)}(p) &=&  \sum_{\mu=1,4} i \ \sin p_{\mu} \cdot \gamma_{\mu} + \kappa ( 1-\cos p_{4}  ) \cdot \gamma_{1} ,\label{eq:mod2-2p}\\
D_{3}^{(2)}(p) &=& \sum_{\mu=1,4}  i \ \sin p_{\mu} \cdot \gamma_{\mu}  + \kappa \sum_{\mu=1,4}( 1-\cos p_{\mu} ) \cdot \gamma_{\mu},\\
D_{4}^{(2)}(p) &=& \sum_{\mu=1,4}  i \ \sin p_{\mu} \cdot \gamma_{\mu}  + \kappa \sum_{\mu,\nu=1,4 \ \mu \ne \nu}( 1-\cos p_{\mu} ) \cdot \gamma_{\nu},\\
D_{5}^{(2)}(p) &=& \sum_{\mu=1,4}  i \ \sin p_{\mu} \cdot \gamma_{\mu}  + \kappa \sum_{\mu=1,4}( 1-\cos p_{\mu} ) \cdot \gamma_{1},
\ee
where $\kappa$ is a hopping parameter which is a real number
\footnote{In this paper, the indices ``1'' and ``4'' denote temporal and spacial components, respectively .}. 
To investigate symmetries of the fermions, we define discrete transformations as
\be
\mathrm{C} &:& \ D(p_{1},p_{4}) \rightarrow  -C D^{\top} (-p_{1},-p_{4}) C^{-1},\nl
\mathrm{P} &:& \ D(p_{1},p_{4}) \rightarrow \gamma_{4} D(-p_{1},p_{4}) \gamma_{4},\\
\mathrm{T} &:& \ D(p_{1},p_{4}) \rightarrow \gamma_{1} D(p_{1},-p_{4}) \gamma_{1},\nn
\ee
where $C$ is a charge conjugation matrix.
In two dimensions, we can choose $C=i \gamma_{1}$ where $\gamma_{1}=\sigma^{2}$ and $\gamma_{4}=\sigma^{1}$.
We also define chiral symmetry and $\gamma_{5}$hermiticity as,
\be
\mathrm{chiral} &:& \ D(p_{1},p_{4})=-\gamma_{5} D(p_{1},p_{4}) \gamma_{5}, \nl
\mathrm{\gamma_{5}hermiticity} &:& \ D(p_{1},p_{4})=\gamma_{5} D^{\dagger}(p_{1},p_{4}) \gamma_{5}, \nn
\ee
where $\gamma_{5}$ is a chiral matrix defined as $\gamma_{5}=i \gamma_{1} \gamma_{4}$.
\begin{table}[t]
  \begin{center} 
  \begin{tabular}{|c||c|c|c|c|c|c|c||c|c|}\hline
 & C & P & T & CP & CT & PT & CPT  & chi & $\gamma_{5}$h \\ \hline \hline
$D_{1}^{(2)}$ & $\times$ & $\times$ & $\bigcirc$ & $\bigcirc$ & $\times$ & $\times$ & $\bigcirc$ & $\bigcirc$ & $\times$  \\\hline 
$D_{2}^{(2)}$ & $\times$ & $\times$ & $\bigcirc$ & $\bigcirc$ & $\times$ & $\times$ & $\bigcirc$ & $\bigcirc$ & $\times$   \\\hline 
$D_{3}^{(2)}$ & $\times$ & $\times$ & $\times$ & $\times$ & $\times$ & $\times$ & $\bigcirc$ & $\bigcirc$ & $\times$   \\\hline 
$D_{4}^{(2)}$ & $\times$ & $\times$ & $\times$ & $\times$ & $\times$ & $\times$ & $\bigcirc$ & $\bigcirc$ & $\times$   \\\hline 
$D_{5}^{(2)}$ & $\times$ & $\times$ & $\bigcirc$ & $\bigcirc$ & $\times$ & $\times$ & $\bigcirc$ & $\bigcirc$ & $\times$  \\\hline 
  \end{tabular}
  \end{center}
  \caption{Discrete symmetries, chiral symmetry and $\gamma_{5}$hermiticity for the Dirac operators $D^{(2)}_{1-5}$.}
  \label{tab:discrete}
\end{table}
All the Dirac operators preserve exact chiral symmetry and CPT symmetry but break C, P, CT, PT and cubic symmetries.
And we can see that $D^{(2)}_{1}$, $D^{(2)}_{2}$ and $D^{(2)}_{5}$ also preserve T and CP symmetries.
We summarize symmetries for the Dirac operators in the Table \ref{tab:discrete}.

\begin{figure}[t]
\begin{minipage}{0.33\hsize}
\begin{center}
\includegraphics[width=50mm]{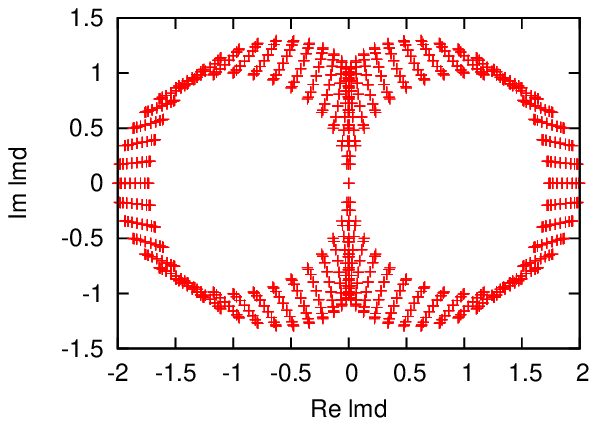}
\hspace{1.6cm}  (a)
\end{center}
\end{minipage}
\begin{minipage}{0.33\hsize}
\begin{center}
\includegraphics[width=50mm]{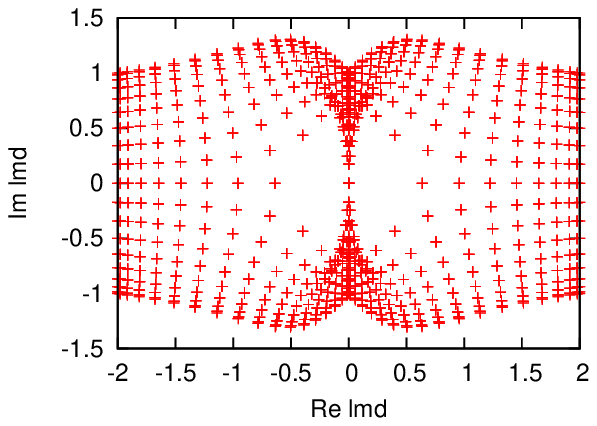}
\hspace{1.6cm}  (b)
\end{center}
\end{minipage}
\begin{minipage}{0.33\hsize}
\begin{center}
\includegraphics[width=50mm]{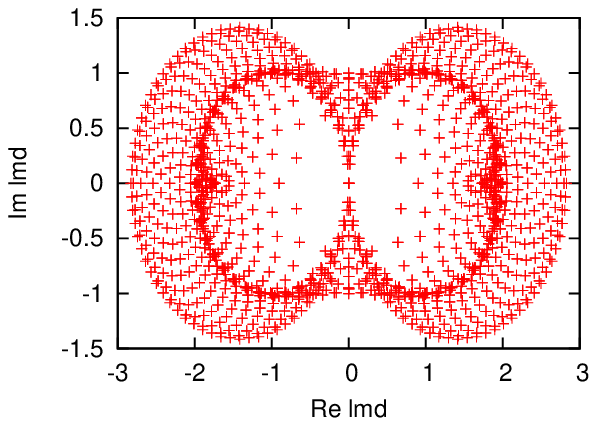}
\hspace{1.6cm}  (c)
\end{center}
\end{minipage}
\begin{minipage}{0.33\hsize}
\begin{center}
\includegraphics[width=50mm]{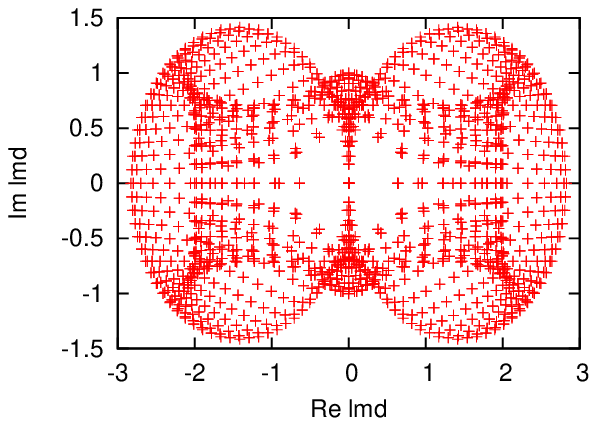}
\hspace{1.6cm}  (d)
\end{center}
\end{minipage}
\begin{minipage}{0.33\hsize}
\begin{center}
\includegraphics[width=50mm]{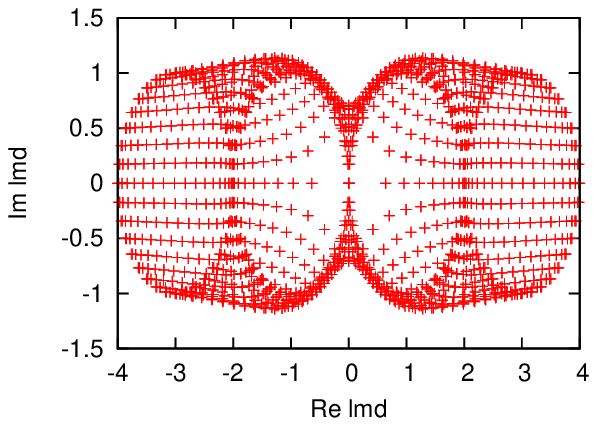}
\hspace{1.6cm}  (e)
\end{center}
\end{minipage}
\caption{The eigenvalue distribution of Dirac operators, (a) $D^{(2)}_{1}$, (b) $D^{(2)}_{2}$, (c) $D^{(2)}_{3}$, (d) $D^{(2)}_{4}$, and (e) $D^{(2)}_{5}$. The horizontal and vertical axes denote the real part and the imaginary part of eigenvalues, respectively. The hopping parameter, the fermion mass and lattice size are fixed at $\kappa=1$, $m=0$ and $L^{2}=36 \times 36$. The blue circle points denote eigenvalues for momenta $p=(0,0)$, $(0,\pi)$, $(\pi,0)$, and $(\pi,\pi)$.}
\label{fig:eigenvalue}
\end{figure}

\begin{figure}[t]
\begin{minipage}{0.33\hsize}
\begin{center}
\includegraphics[width=50mm,angle=0]{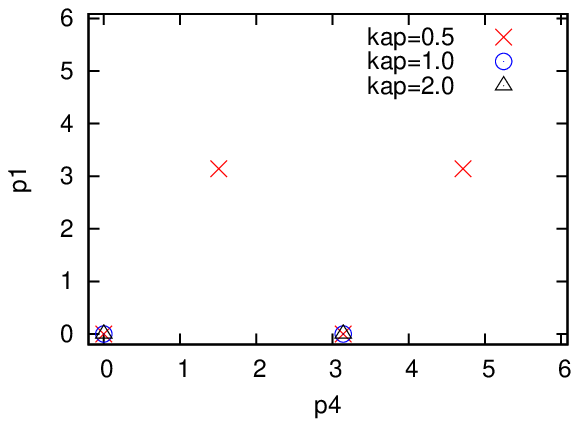}
\hspace{1.6cm}  (a)
\end{center}
\end{minipage}
\begin{minipage}{0.33\hsize}
\begin{center}
\includegraphics[width=50mm,angle=0]{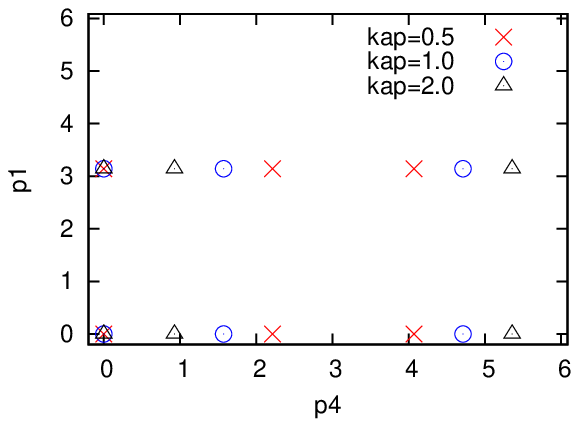}
\hspace{1.6cm}  (b)
\end{center}
\end{minipage}
\begin{minipage}{0.33\hsize}
\begin{center}
\includegraphics[width=50mm,angle=0]{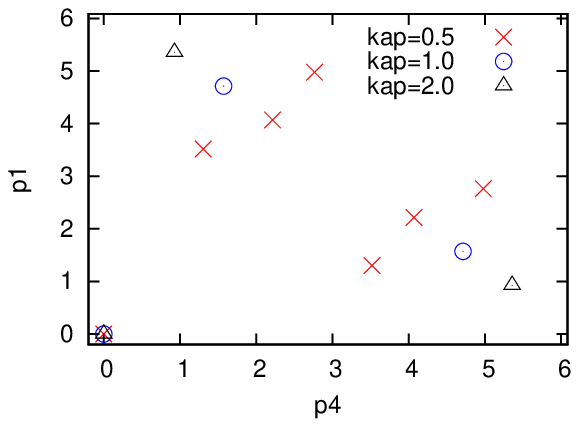}
\hspace{1.6cm}  (c)
\end{center}
\end{minipage}
\begin{minipage}{0.33\hsize}
\begin{center}
\includegraphics[width=50mm,angle=0]{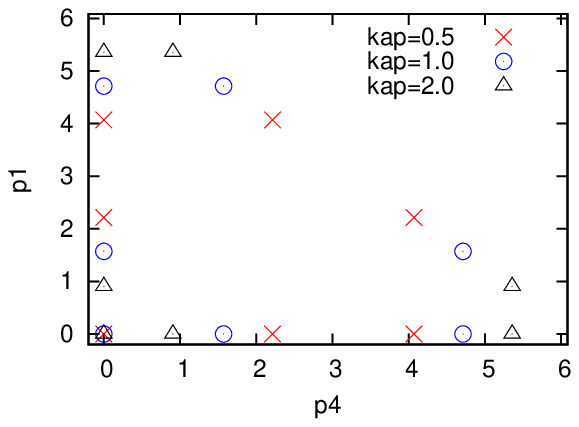}
\hspace{1.6cm}  (d)
\end{center}
\end{minipage}
\begin{minipage}{0.33\hsize}
\begin{center}
\includegraphics[width=50mm,angle=0]{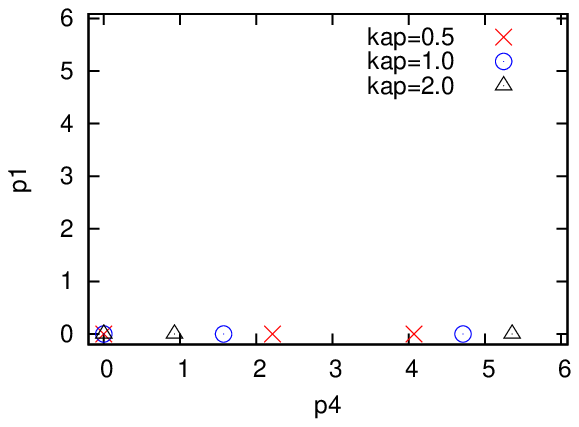}
\hspace{1.6cm}  (e)
\end{center}
\end{minipage}
\caption{The poles of five Dirac operators, (a) $D^{(2)}_{1}$, (b) $D^{(2)}_{2}$, (c) $D^{(2)}_{3}$, (d) $D^{(2)}_{4}$, and (e) $D^{(2)}_{5}$. The horizontal and vertical axes denote momenta $p_{4}$ and $p_{1}$, respectively. The cross points, circle points, and triangle points represent poles at $\kappa=0.5, 1$, and $2$, respectively.}
\label{fig:pole}
\end{figure}

To understand more about the fermions, we examine the eigenvalue distributions and the number of poles.
The Fig.\ref{fig:eigenvalue} shows eigenvalue distribution of the fermions.
We clearly see that the eigenvalues of the $D^{(2)}_{2}$, $D^{(2)}_{3}$, $D^{(2)}_{4}$ and $D^{(2)}_{5}$ spread over the $\mathrm{Re}\lambda$-$\mathrm{Im}\lambda$ plane entirely in the continuum limit.
On the other hand, in the $D^{(2)}_{1}$ case, the eigenvalues localize along the $\mathrm{Im}\lambda$ axis in the limit.
Fig.\ref{fig:pole} represents the number of poles of the fermions at $\kappa=0.5,1$ and $2$.
The $D^{(2)}_{1}$ at $\kappa \ne 1$ generates $p=(0,0)$ and $(0,\pi)$.
However, the other fermions have more than two or odd number of poles .

Using appropriate regularization, such as small mass and anti-boundary condition, the Dirac determinant of $D^{(2)}_{2-5}$ can be evaluated.
However, in the principle of lattice gauge theory, any lattice fermion must return the continuum fermion in the continuum limit.
The $D^{(2)}_{2-5}$ return the continuum classical action, but the eigenvalues of the Dirac operators do not recover in the limit.
Therefore, the Dirac operators should be excluded from a candidate for lattice theory.


\section{Gross-Neveu model in two dimensions} \label{sec:aoki}
We apply the non-$\gamma_{5}$hermiticity fermion to the two dimensional Gross-Neveu model.
The model can be solved exactly in large-$N$ limit using the saddle point approximation.
In this limit, the parity broken phase diagram, called Aoki phase\cite{Aoki:1983qi}, can be obtained.
We also focus on the model with an imaginary chemical potential and the chiral symmetry broken phase diagram \cite{Misumi:2012uu}.
In this work, we apply only the Dirac operator $D^{(2)}_{1}$ defined in Eq.(\ref{eq:mod2-1}).
At a first sight, the fermion seems to be improper for a practical calculation because it does not preserve $\gamma_{5}$hermiticity.
The analyzes mentioned below are verification for application of the fermion to simple cases.

We define the lattice action of the Gross-Neveu model as
\be
S_{\mathrm{lat,aux}}^{\mathrm{GN}} &=& \sum_{n,m}  \bar{\psi}_{n} \left[ \ \bar{D}^{(2)}_{1 \ nm} + \left( m  + \sigma_{n} + \pi_{n} i \gamma_{5}  \ \right) \delta_{n,m}  \right] \psi_{m}  + \frac{N}{2} \sum_{n} \left[ \frac{\sigma^{2}_{n}}{g^{2}_{\sigma}}  + \frac{\pi^{2}_{n}}{g^{2}_{\pi}}  \right] ,  \\ \label{eq:GNlaux}
S_{\mathrm{lat,aux}}^{\mathrm{GNim}} &=& \sum_{n,m}  \bar{\psi}_{n} \left[ \ D^{(2)}_{1 \ nm} + \left(   \sigma_{n} + \pi_{4 n} i \gamma_{4}  \ \right) \delta_{n,m}  \right] \psi_{m} + \frac{N}{2} \sum_{n} \left[ \frac{1}{g^{2}_{\sigma}} (\sigma_{n}-m)^{2} + \frac{1}{g^{2}_{\pi}} (\pi_{4 n}-\mu)^{2}  \right] , \label{eq:GNlchem} \nl
\ee
where $\mu$ is a chemical potential, and $\sigma, \pi$, and $\pi_{4}$ are an auxiliary scalar field, an auxiliary pseudo-scalar, and an pseudo-vector field, respectively.
The lower indices in these actions denote a coordinate in the lattice space.
We omit the flavor indices, $\bar{\psi}\psi \equiv \sum_{i=1}^{N} \bar{\psi}^{i} \psi^{i}$.
We can obtain original actions including four-fermi couplings by integrating out the auxiliary fields $\sigma$, $\pi$, and $\pi_{4}$.
These actions preserve not only $\mathrm{U_{V}(1)}$ symmetry but also chiral ${\bf Z}_{4}$ symmetry in the massless case.
In the (\ref{eq:GNlaux}) case, chiral ${\bf Z}_{4}$ symmetry is enhanced to ${\rm U_{A}(1)}$ symmetry at $g^{2}_{\sigma}=g^{2}_{\pi}$.

Now, we choose two Dirac operators for the Gross-Neveu model with or without an imaginary chemical potential as follows: 
\be
\bar{D}^{(2)}_{1\ nm} &=& \sum_{\mu=1,4} \frac{1}{2} \left( \delta_{n+\hat{\mu},m} - \delta_{n-\hat{\mu},m} \right) \cdot \gamma_{\mu} +\frac{\kappa}{2} \left(2 \delta_{n,m}-\delta_{n+\hat{4},m} - \delta_{n-\hat{4},m} \right) \cdot \gamma_{4}, \label{eq:mod-br}  \\
D^{(2)}_{1 \ nm} &=& \sum_{\mu=1,4} \frac{1}{2} \left( \delta_{n+\hat{\mu},m} - \delta_{n-\hat{\mu},m} \right) \cdot \gamma_{\mu} +\frac{\kappa}{2} \left(2 \delta_{n,m}-\delta_{n+\hat{1},m} - \delta_{n-\hat{1},m} \right) \cdot \gamma_{1},  \label{eq:mod2-1}
\ee
In the large N limit, we can derive the following gap equations for the critical line of Aoki phase using the saddle point approximation, 
\be
\frac{\sigma_{0}}{g^{2}_{\sigma}} =  \int \frac{d^{2}k}{(2 \pi)^{2}} \frac{2 \sigma_{m_{\mathrm{c}}}}{\sigma_{m_{\mathrm{c}}}^{2} +  H(k)},\qquad
 \frac{1}{g^{2}_{\pi}} =  \int \frac{d^{2}k}{(2 \pi)^{2}} \frac{2}{\sigma_{m_{\mathrm{c}}}^{2} + H(k)} \label{eq:cri-line1},
\ee
where
\be
H(k) &=& \sin^{2} k_{1} +\left[ \sin k_{4}-i \kappa(1-\cos k_{4}) \right]^{2}, 
\ee
$m_{\mathrm{c}}$ is a critical mass, and $\sigma_{m_{\mathrm{c}}}=m_{\mathrm{c}}+\sigma_{0}$. 
And the lower index $0$ denotes the solutions for the saddle point approximation.
In the similar way, gap equations for the chiral broken phase are obtained as,
\be
\frac{1}{g^{2}_{\sigma}} =   \int \frac{d^{2}k}{(2 \pi)^{2}} \frac{2}{ \tilde{H}(k)}, \qquad
 \frac{\pi_{40}-\mu_{c}}{g^{2}_{\pi}} = \int \frac{d^{2}k}{(2 \pi)^{2}} \frac{2(\pi_{40}+\sin k_{4})}{ \tilde{H}(k)},
\ee
where 
\be
\tilde{H}(k) &=& \left[ \sin k_{1}-i \kappa(1-\cos k_{1}) \right]^{2}
              +(\sin k_{4}+\pi_{4 0})^{2},
\ee
and $\mu_{c}$ is a critical chemical potential.
We also obtain the Aoki phase of the fermion adding flavored mass \cite{Creutz:2011cd} defined as
\be
m &\rightarrow& m +m_{f}(p), \nl
m_{f}(p) &=&
\begin{cases}
m_{f} \cdot \cos p_{1} \cos p_{4} &  \mbox{for the naive fermion} \\
m_{f} \cdot \cos p_{1} &  \mbox{for the non- $\gamma_{5}$hermiticity fermion} 
\end{cases},
\ee
where the $m_{f}$ in the r.h.s is a constant.

\begin{figure}[t]
\begin{minipage}{1.0\hsize}
\begin{center}
\includegraphics[width=70mm]{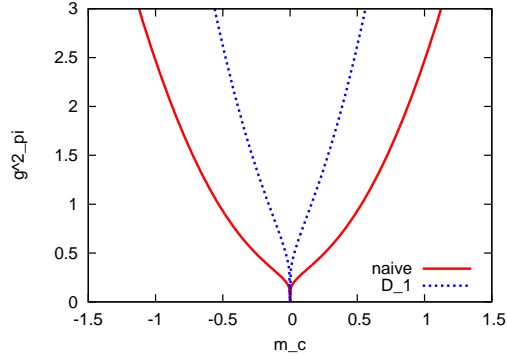}
\end{center}
\end{minipage}
\caption{The Aoki phase using the naive fermion (Red line) and the non-$\gamma_{5}$hermiticity fermion $\bar{D}^{(2)}_{1}$ (Blue line) at $g^{2}_{\sigma}=g^{2}_{\pi}/2$.
We fixed the hopping parameter of the non-$\gamma_{5}$hermiticity fermion at $\kappa=1$.
The horizontal and vertical axes denote the critical mass $m_{c}$ and the squared four-fermi coupling constant $g^{2}_{\pi}$, respectively.}
\label{fig:aoki}
\end{figure}

\begin{figure}[t]
\begin{minipage}{0.5\hsize}
\begin{center}
\includegraphics[width=70mm]{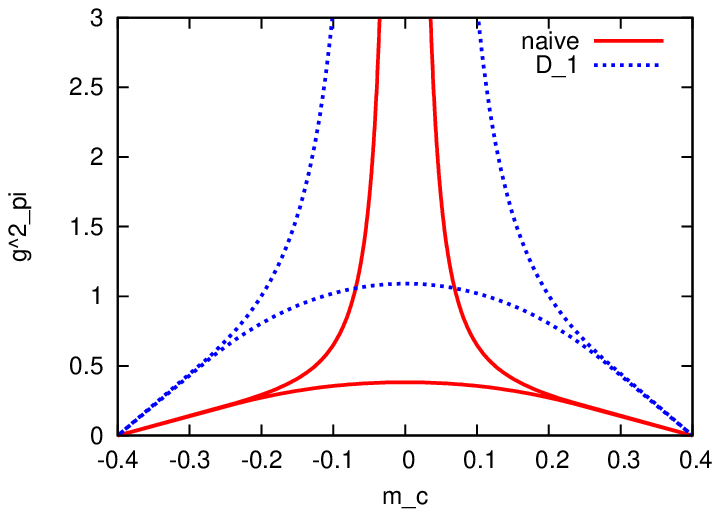}
\end{center}
\end{minipage}
\begin{minipage}{0.5\hsize}
\begin{center}
\includegraphics[width=70mm]{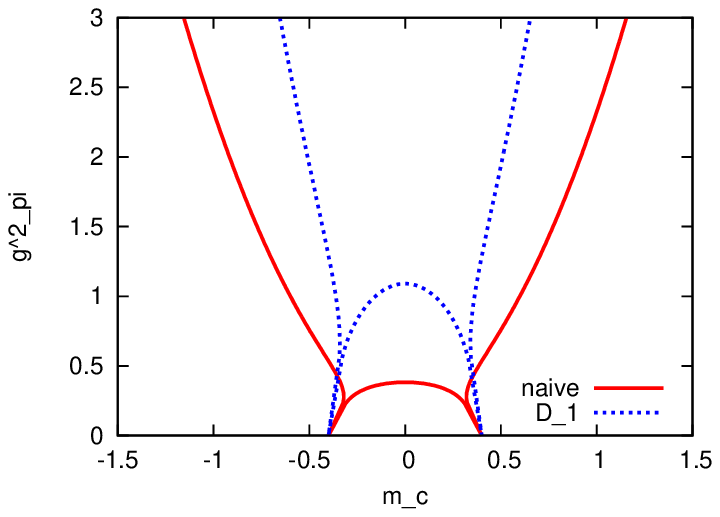}
\end{center}
\end{minipage}
\caption{The Aoki phase using the naive fermion (Red line) and the non-$\gamma_{5}$hermiticity fermion $\bar{D}^{(2)}_{1}$ (Blue line) adding the flavored mass term at (Left) $g^{2}_{\sigma}=g^{2}_{\pi}$ and (Right) $g^{2}_{\sigma}=g^{2}_{\pi}/2$.
We fix the factor of flavored mass and the hopping parameter of the non-$\gamma_{5}$hermiticity fermion at $m_{f}=0.4$ and $\kappa=1$, respectively.
The horizontal and vertical axes denote the critical mass $m_{c}$ and the squared four-fermi coupling constant $g^{2}_{\pi}$, respectively.}
\label{fig:aoki-fm}
\end{figure}

\begin{figure}[ht]
\begin{minipage}{0.5\hsize}
\begin{center}
\includegraphics[width=70mm]{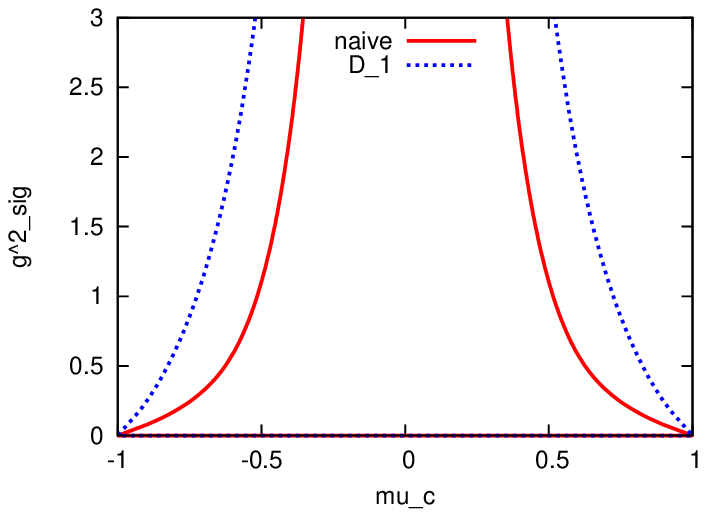}
\end{center}
\end{minipage}
\begin{minipage}{0.5\hsize}
\begin{center}
\includegraphics[width=70mm]{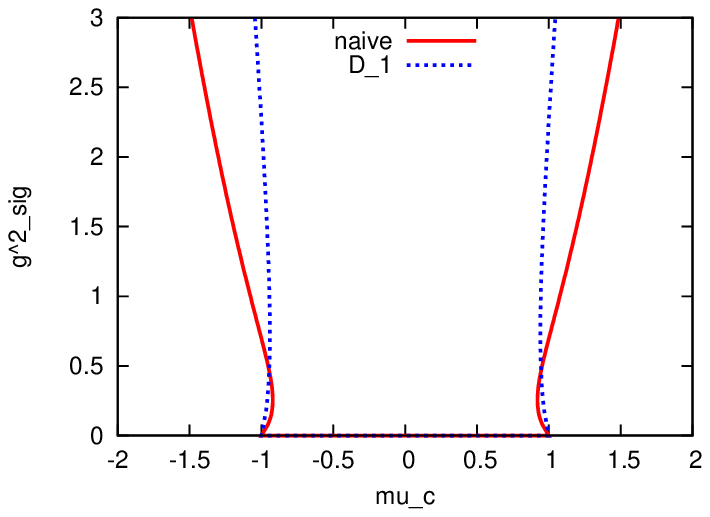}
\end{center}
\end{minipage}
\caption{The chiral broken phase diagrams using the naive fermion (Red line) and the non-$\gamma_{5}$hermiticity fermion $D^{(2)}_{1}$ (Blue line) at (Left) $g^{2}_{\sigma}=g^{2}_{\pi}$ and (Right) $g^{2}_{\sigma}=2g^{2}_{\pi}$.
We fix the hopping parameter in the non-$\gamma_{5}$hermiticity fermion at $\kappa=2$.
The horizontal and vertical axes denote the critical chemical potential and the squared four-fermi coupling constant $g^{2}_{\sigma}$, respectively.}
\label{fig:im_chem}
\end{figure}

The results are in Fig.\ref{fig:aoki}, Fig.\ref{fig:aoki-fm}, and Fig.\ref{fig:im_chem}
\footnote{In the phase diagrams using the fermions adding the flavored mass, there is first-order phase transition at the bottom of the diagrams enclosed the critical line, so that the gap equations defined in Eqs.(\ref{eq:cri-line1}), are not applicable in this region. 
However, we take leave not to correct to emphasize that we can obtain solutions for the gap equations.}.
We can see that the phase structures using the non-$\gamma_{5}$hermiticity fermion are very similar to the naive fermion cases. 
All the couping constants are real numbers despite of using fermions without $\gamma_{5}$hermiticity.



\section{Summary and discussion} \label{sec:sum}
We formulated new two dimensional fermions without $\gamma_{5}$hermiticity (non-$\gamma_{5}$hermiticity fermion), based on the minimal doubling fermion, and investigated some properties of the fermions.
The fermions preserve translation invariance, exact chiral symmetry and locality, but break cubic symmetry and some discrete symmetries.
The eigenvalues of $D^{(2)}_{1}$, defined in Eq.(\ref{eq:mod2-1p}), distributes along the imaginary axis in the continuum limit, but the others do not.
In general, the fermions either have at least four poles, or else an odd number poles, but $D^{(2)}_{1}$  at $\kappa \ge 1$ generates two poles.

We also applied the fermion, $D^{(2)}_{1}$, to the two dimensional Gross-Neveu model with or without imaginary chemical potential.
In the application, the parity broken phase diagram which is called Aoki phase and the chiral symmetry broken phase diagram were obtained.
The phase structures are similar to the ones obtained with the naive fermion, and all the coupling constants are real numbers despite of $\gamma_{5}$hermiticity breaking.

In the continuum theory, the eigenvalues of Dirac operator without $\gamma_{5}$hermiticity spread over the whole ${\rm Re} \lambda$-${\rm Im} \lambda$ plane.
Therefore, the behavior of eigenvalues of $D^{(2)}_{1}$ can be interpreted as an effect of the lattice spacing.
In the case of the non-$\gamma_{5}$hermiticity fermions, each spectrum is paired with complex conjugate, so that the Dirac determinant is real-valued.

We will investigate gauge theory using the non-$\gamma_{5}$hermiticity fermion and higher dimensional extension in future work.

\end{document}